\documentclass[a4paper, conference, 10pt] {IEEEtran}
\usepackage{amsmath,amssymb}
\usepackage{amsthm}
\usepackage[pdftex]{graphicx}
\usepackage{tabularx}
\usepackage{tikz}
\definecolor{cof}{RGB}{219,144,71}
\definecolor{pur}{RGB}{186,146,162}
\definecolor{greeo}{RGB}{91,173,69}
\definecolor{greet}{RGB}{52,111,72}
\usetikzlibrary{patterns}
\usetikzlibrary{arrows}
\usepackage{pgfplots}
\usepackage{hhline}
\usepackage{enumerate}
\begin{document}
\title{Minimizing Age of Information in a Multihop Wireless Network}
\author{\IEEEauthorblockN{Ashok Krishnan K.S. and Vinod Sharma}
	\IEEEauthorblockA{Dept. of ECE, Indian Institute of Science, Bangalore, India\\
		Email: \{ashokk, vinod\}@iisc.ac.in
	}}

\maketitle

\begin{abstract}
	We consider the problem of minimizing age in a multihop wireless network. There are multiple source-destination pairs, transmitting data through multiple wireless channels, over multiple hops. We propose a network control policy which consists of a distributed scheduling algorithm, utilizing channel state information and queue lengths at each link, in combination with a packet dropping rule. Dropping of older packets locally at queues is seen to reduce the average age of flows, even below what can be achieved by Last Come First Served (LCFS) scheduling. Dropping of older packets also allows us to use the network without congestion, irrespective of the rate at which updates are generated. Furthermore, exploiting system state information substantially improves performance. The proposed scheduling policy obtains average age values close to a theoretical lower bound as well.
\end{abstract}

\begin{IEEEkeywords}
	age of information, scheduling, multihop networks.
\end{IEEEkeywords}
\section{Introduction}
Consider a monitoring system that observes a physical process, and sends its observations to another location, over a wireless network. Such systems arise naturally in various contexts, and are set to become even more prominent in the context of the Internet of Things (IoT) \cite{IoTSurvey}. These include health monitoring systems, security devices and safety monitors. In many of these applications, it is crucial that the data received at the destination is \emph{fresh}. More recently generated data is considered fresher than data generated earlier. Age of Information (AoI) is a recently introduced metric, that quantifies the freshness (or staleness) of information in such a communication system. It has received considerable interest owing to the fact that there are a number of applications which require fresh information to be delivered from one point to another; the relevance of packets that are generated earlier decays over time. 

\begin{figure}
 	\centering
 	\setlength{\unitlength}{1cm}
 	\thicklines
 	\begin{tikzpicture}[scale=0.7, transform shape]
 	\node (v1) at (1,-0.5) {\huge $t_1$};
	\node (v2) at (4,-0.5) {\huge $t_2$};
	\node (v3) at (3,-0.5) {\huge $\hat t_1$};
	\node (v4) at (6.5,-0.5) {\huge $\hat t_2$};
	\node (v5) at (-0.8,3) {\huge $\alpha(t)$};
	\node (v6) at (8,-0.5) {\huge $t$};
	\draw[->,line width=0.2mm] (-0.8,4) -- (-0.8,5);
	\draw[->,line width=0.2mm] (8.5,-0.5) -- (9.5,-0.5);
	\draw[->,line width=0.4mm] (0,0) -- (0,6);
	\draw[->,line width=0.4mm] (0,0) -- (10,0);
	\draw[line width=0.4mm,color=red,dashed] (0,0) -- (3,3);
	\draw[line width=0.4mm,color=blue,dashed] (0,0) -- (1,1);
	\draw[line width=0.4mm,color=blue,dashed] (1,1) -- (1,0);
	\draw[line width=0.4mm,color=blue,dashed] (1,0) -- (4,3);
	\draw[line width=0.4mm,color=blue,dashed] (4,3) -- (4,0);
	\draw[line width=0.4mm,color=blue,dashed] (4,0) -- (8,4);
	\draw[line width=0.4mm,color=red,dashed] (3,3) -- (3,2);
	\draw[line width=0.4mm,color=red,dashed] (3,2) -- (6.5,5.5);
	\draw[line width=0.4mm,color=red,dashed] (6.5,5.5) -- (6.5,2.5);
	\draw[line width=0.4mm,color=red,dashed] (6.5,2.5) -- (8.5,4.5);
	\draw[line width=0.4mm,dotted] (3,0) -- (3,2);
	\draw[line width=0.4mm,dotted] (6.5,0) -- (6.5,2.5);
	\end{tikzpicture}
 	\caption{Evolution of Age. The red and blue lines show the evolution of the age of information at the destination and source respectively, as a function of time. At times $t_1$ and $t_2$, the first and second packets are generated at the source. These are received at the destination at times $\hat t_1$ and $\hat t_2$.}
 	\label{figAgeEvol}
 \end{figure}
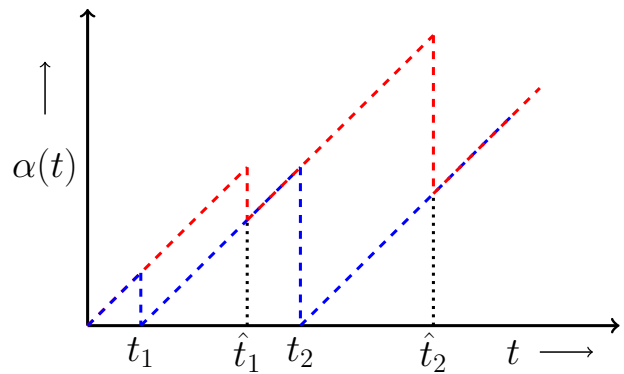

The notion of AoI \cite{Yates, kaul2011minimizing} enabled a new understanding of freshness of information. Consider a source generating packets  to be sent to a destination, across a network. Let the packets be generated at the source at times $t_1,t_2,t_3,\dots$, and received at the destination at  $\hat t_1,\hat t_2,\hat t_3\dots$ (the packets need not be received in the same order in which they were generated). Define,
\begin{align}
n^*(t)=\arg_n\max\{t_n:\hat t_n\leq t\}.
\end{align}
This is the index of that packet among all packets received at the destination, till time $t$, which has been generated most recently, i.e., the freshest packet present at the destination. The age of information is defined as the age of this packet, i.e.,
\begin{align}\label{ageDefn}
\alpha(t)=t-t_{n^*(t)}.
\end{align}
The evolution of the age function $\alpha(t)$ is given in Figure \ref{figAgeEvol}. Note that AoI can be defined for the source as well, seeing it as a point that receives the packets with zero delay.
Define the Average AoI $\bar\alpha(t)$ as,
\begin{align}\label{avgAgeDefn}
\bar\alpha(t)=\frac{1}{t}\int_0^t\alpha(\tau)d\tau.
\end{align}
We will refer to the (average) age at the destination node to be the (average) age of the flow.  Between the source and destination, packets experience queueing delays and transmission delays. While queueing delay contributes to the age process, delay is not identical to age. The age process depends on both the queueing delay and the rate at which packetized updates are being generated at the source. One can reduce the packet generation rate, which may lead to lower buffer levels, and hence, lower delays. However, owing to fewer updates, the age process may not reduce. On the other hand, sending too many updates may lead to congestion in the network.  

The earliest works in AoI literature model the source to destination transmission system as a single hop queue. In \cite{Yates}, the problem of minimizing the average AoI for $M/M/1$,  $D/M/1$ and $M/D/1$ queues, under the First Come First Served (FCFS) discipline, is studied. Analytical expressions were obtained for Average AoI in the first two cases, and it was seen that there was an optimal load factor at which Average AoI was minimized. However, obtaining explicit expressions for AoI may not be easy under other service disciplines or complex network assumptions. Later works looked at AoI for other single queue models, such as sharing of an $M/M/1$ FCFS queue by two traffic streams \cite{YatesKaul2017},  an $M/M/1$ Last Come First Served (LCFS) queueing system  with and without preemption \cite{YatesKaul2016}, and an $M/M/2$ system \cite{Ephrem}. In \cite{Kadota}, the authors consider a single base station, with multiple nodes trying to communicate time sensitive data to it, and propose three policies to minimize average AoI subject to throughput requirements. They also show that the AoI obtained in their policies is a multiplicative factor away from the optimal value. In \cite{najm2018content} the authors study the problem of giving pre-emptive priority to one flow over another, in a single queue system, and obtain closed-form expressions for average age and peak age. In works such as \cite{sun2018sampling,bacinoglu2015age}, the problem of optimal sampling in order to minimize age is addressed, in the context of single hop transmission. However, they do not take into account the effects on queueing owing to a higher sampling rate.

More generally, one may model the source to destination transmission system as a multihop network. This models transmission of observations across a network which could be local, or even the internet.  In the case of multihop networks, there have been a number of studies of the AoI problem.  In \cite{Shroff}, the authors consider a multihop network with a single flow. Under the assumption that service times are  exponentially distributed, they show that the (preemptive) Last Come First Served (LCFS) service discipline minimizes the age among all disciplines, in a stochastic ordering sense. In \cite{ModianoDist}, the authors study distributed stationary policies that are not dependent on the channel state. Using these policies, they obtain tractable expressions for Average and Peak AoI, which are then optimized over this class of policies. However, this class of policies may be a small subset of all possible policies, and  therefore not very likely to contain the policy that minimizes age among all possible policies. In \cite{EryilmazHeavy}, the authors propose an age based maxweight type scheduling policy that is throughput optimal, and further provide heavy traffic approximations for its performance. A concise survey covering diverse aspects of AoI, and giving a number of available AoI results for different system models, is  \cite{Kosta}.

In this work we look at the AoI problem for a multihop network with multiple flows. The contributions of this paper are summarized below.
\begin{itemize}
\item We present the State Dependent Scheduling with Packet Dropping policy (SDSPD). The system state consists of the queue lengths at different nodes, channel gains and the age of each flow at its destination. SDSPD provides a scheduling rule and a service discipline. The service rule consists of dropping older packets at each queue.
\item The SDSPD policy results in an age at least as low as that achieved by LCFS. Due to reduction of number of packets in the system, we are actually able to perform better, as demonstrated by simulations. We also compare the ages obtained with a theoretical lower bound, and show that the system performs close to this.
\item Due to the packet dropping rule, we are ensured of stability at all arrival rates. The system can accommodate samples arriving at any rate, without congestion. From simulations, it can be seen that it outperforms policies that do not drop packets, as well as stationary policies which do not take into account the system state (queue lengths and channel gains) while making control decisions.
\item Packet dropping also enables us to increase arrival rates without leading to congestion. Age is seen to reduce with arrival rate. Thus, network capacity is not a constraint on the sampling rate, and can be optimized independently.
\item Using an optimization with dynamically varying weights, we provide average age close to the desired targets. We provide a distributed version of the algorithm as well.
\end{itemize}
The remainder of this paper is organized as follows. In Section II, we provide the system model, formulate the problem and propose a control policy. The performance of the policy is analysed by simulations and compared with existing policies. The results obtained are discussed in Section III. Subsequently we have the concluding remarks in Section IV.

\section{System Model and Problem Formulation}
 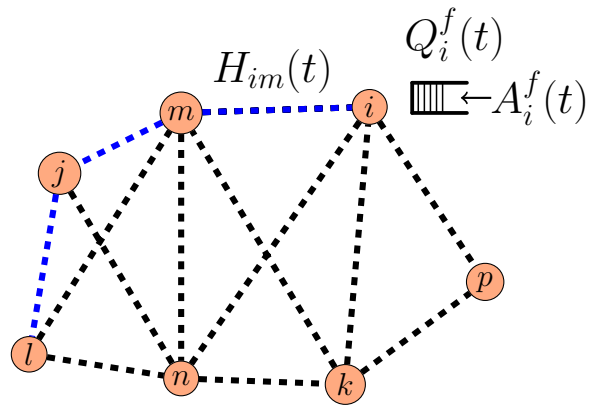
\begin{figure}
 	\centering
 	\setlength{\unitlength}{1cm}
 	\thicklines
 	\begin{tikzpicture}[scale=0.8, transform shape]		
 	\node[draw,shape=circle, fill={rgb:orange,1;yellow,0;pink,2;green,0}, scale=0.6, transform shape] (v1) at (4.7,0.5) {\Huge $k$};
 	\node[draw,shape=circle, fill={rgb:orange,1;yellow,0;pink,2;green,0}, scale=0.6, transform shape] (v2) at (2.0,0.6) {\Huge$n$};
 	\node[draw,shape=circle, fill={rgb:orange,1;yellow,0;pink,2;green,0}, scale=0.6, transform shape] (v3) at (2.0,5.0) {\Huge$m$};
 	\node[draw,shape=circle, fill={rgb:orange,1;yellow,0;pink,2;green,0}, scale=0.6, transform shape] (v4) at (0.0,4) {\Huge$j$};
 	\node[draw,shape=circle, fill={rgb:orange,1;yellow,0;pink,2;green,0}, scale=0.6, transform shape] (v5) at (5.1,5.1) {\Huge$i$};
 	\node[draw,shape=circle, fill={rgb:orange,1;yellow,0;pink,2;green,0}, scale=0.6, transform shape] (v6) at (7,2.2) {\Huge$p$};
 	\node[draw,shape=circle, fill={rgb:orange,1;yellow,0;pink,2;green,0}, scale=0.6, transform shape] (v10) at (-0.5,1.0) {\Huge$l$};
 	\node (v7) at (7.9,5.25) {\huge $A_i^f(t)$};
 	\node (v8) at (6.5,6.3) {\huge $Q_i^f(t)$};
 	\node (v9) at (3.5,5.7) {\huge $H_{im}(t)$};			
 	\draw[line width=0.8mm, dashed] (v2) -- (v1)
 	(v4) -- (v2)
 	(v2) -- (v5)
 	(v3) -- (v5)
 	(v3) -- (v2)
 	(v1) -- (v5)
 	(v1) -- (v3)
 	(v3) -- (v5)
 	(v5) -- (v6)
 	(v10) -- (v2)
 	(v10) -- (v3)
 	(v1) -- (v6);
 	\draw[line width=0.8mm, dashed, color=blue] (v3) -- (v5) (v3) -- (v4) (v4) -- (v10);
 	\draw[line width=0.5mm, line cap=round](5.8,5)--(6.7,5);
 	\draw[line width=0.5mm, line cap=round](5.8,5.5)--(6.7,5.5);
 	\draw[line width=0.5mm, line cap=round](5.8,5)--(5.8,5.5);
 	\draw[line width=0.2mm](5.9,5)--(5.9,5.5);
 	\draw[line width=0.2mm](6.0,5)--(6.0,5.5);
 	\draw[line width=0.2mm](6.1,5)--(6.1,5.5);
 	\draw[line width=0.2mm](6.2,5)--(6.2,5.5);
 	\draw[line width=0.2mm](6.3,5)--(6.3,5.5);
 	\draw[thick,->] (7.1,5.25) -- (6.6,5.25);
 	\end{tikzpicture}
 	\caption{A simplified depiction of a Wireless Multihop Network. The flow $f$ follows the path $i\to m\to j\to l$.}
 	\label{figSysModel}
 \end{figure}
We consider a multihop wireless network (see Fig (\ref{figSysModel})), modelled as a graph $\mathcal G=(\mathcal V,\mathcal E)$, where $\mathcal V$ is the set of nodes, and $\mathcal E\subseteq\mathcal V\times\mathcal V$ is the set of edges (links) on $\mathcal V$. Packets are generated at \emph{source nodes}, to be sent to various \emph{destination nodes}. Each such stream of packets, corresponding to a source-destination pair, is called a \emph{flow}. The set of all flows in the network will be denoted by $\mathcal F$. For any flow $f$, we use $src(f)$ and $des(f)$ to denote its source and destination nodes. For each flow, a \emph{path} is a set of nodes connecting the source to its destination. We assume that paths are fixed and known a priori. This would imply that a routing algorithm was employed beforehand to create these routes (see \cite{akkaya2005survey} for a survey of common routing algorithms in wireless sensor networks).

We have a slotted system, with time index $t\in\{0,1,2,\dots\}$. Each slot is of unit length and time duration $[t,t+1)$ denotes slot $t$. The arrival process for a flow $f$ with source node $i$ is denoted by $A_i^f(t)$. We assume that $A_i^f(t)$ evolves as an  independent and identically distributed (i.i.d.) sequence across time slots and independent of other flows. The wireless channel gain of a link $(i,j)\in\mathcal E$ at time $t$ will be denoted by $H_{ij}(t)$. This is also i.i.d. across time for a link, and is independent across links. The overall channel state is denoted by $H(t)=\{H_{ij}(t)\}_{(i,j)\in\mathcal E}$. We transmit at a constant power and a fixed rate. If a channel gain  is above a threshold and interference from other channels is limited then we assume that there  is a successful transmission. At each node $i$, there is a queue $Q_i^f(t)$ which consists of packets of flow $f$ present at node $i$. Let $S_{ij}^f(t)$ denote the number of packets of flow $f$ sent over link $(i,j)$ in time slot $t$. The queues evolve as,
\begin{align}
Q_i^f(t+1)=Q_i^f(t)+\sum_jS_{ji}^f(t)-\sum_kS_{ik}^f(t),
\end{align}
where $i\neq des(f)$. By $\alpha^f(t)$ and $\bar\alpha^f(t)$ we denote the AoI and average AoI of flow $f$ at its destination (see (\ref{ageDefn}), (\ref{avgAgeDefn})). 

We assume that the links fall into interference sets. An interference set is a subset of $\mathcal E$ such that no two members of that set can transmit simultaneously. These define the \emph{interference constraints} of the system. Subject to these constraints, only certain configurations of links can be activated at a time. 

A \emph{schedule} $s$ is a mapping $s:\mathcal E\times\mathcal F\to\{0,1\} $. If $s(e,f)=1$, then flow $f$ is scheduled to be transmitted on link $e$ in that slot. Not all mappings from $\mathcal E\times\mathcal F$ to $\{0,1\}$ are feasible schedules. The links that are active must obey the interference constraints. Further, two flows cannot be simultaneously scheduled on a link. The schedules that obey these constraints are called \emph{feasible} schedules. Denote the set of all feasible schedules by $\mathcal S$. Corresponding to each feasible schedule $s$ and channel state $H$, there is a rate vector $R=\{R_{ij}^f\}_{(i,j)\in\mathcal E,f\in\mathcal F}$. We are interested in obtaining control policies that can reduce the AoI at the destinations. To this end, we propose the following policy.
\subsection{Control Policy}
The control policy we propose is called State Dependent Scheduling with Packet Dropping (SDSPD). This policy consists of a service discipline and an optimization rule.
\subsubsection{Service Discipline}
Under the SDSPD policy, at each queue, we keep only the latest packet of a flow, and all others are discarded. Thus, if a more recently generated packet of a flow is received at a queue, all packets generated prior to that packet of that flow at the queue are dropped. This is a local decision that can be implemented at the node level. There is no need for exchange of information between the nodes for this purpose. Consequently, at all nodes $i$ and for all flows $f$, $Q_i^f\in\{0,1\}$. Such a service discipline will result in a performance similar to (or better than) an LCFS discipline.
\subsubsection{Optimization Rule}
The schedule at time $t$ is chosen to be $s^*(t)$, where,
\begin{align}\label{OptRule}
s^*(t)=\arg_{s\in\mathcal S}\max\sum_{i,j,f}w^f(\alpha^f(t))Q_i^f(t)R_{ij}^f(s,H(t)),
\end{align}
where $w^f$ is the weight for flow $f$, which is a function of the age $\alpha^f(t)$ of flow $f$ at time $t$ at its destination node. Also,
\begin{equation}\label{weightDefn}
w^f(x)=
\begin{cases}
& 1\ \text{if}\ x<\bar\alpha^f,\\
& 1+\beta\ \text{if}\ x\geq\bar\alpha^f,
\end{cases}
\end{equation}
where $\bar\alpha^f$ is a desired average age for flow $f$, and $\beta$ is a fixed positive quantity. This represents a weighted queue policy with dynamic weights. The weight function $w^f$ enables us to differentiate between the flows, and gives higher priority to some flows, if desired. A flow with a higher weight will be scheduled more often, and consequently its age should decrease. A lower $\bar{\alpha}^f$ gives higher priority to flow $f$.\\
\indent Note that the quantity being optimized is different from the traditional maxweight metric, which involves a \emph{backpressure} term.  Owing to the packet dropping in our system, the vector $Q(t)$ remains in a bounded set for all time $t$, and consequently, the system is always stable. Hence, we do not use a maxweight formulation, which is used generally to guarantee stability (within the capacity region of the system).  

We will see in Section III that this policy is seen to yield a good performance in terms of the average AoI metric. We compare it with multiple policies, and see the benefit of dropping packets, even  compared to policies which do LCFS. In the following section, we describe how we may solve the optimization problem in a distributed manner.
\subsection{Distributed Implementation}
While the optimization (\ref{OptRule}) may be non-convex in general, in case of smaller state spaces, it can be computed by a brute force search. For larger state spaces, it can be approximated by a linear relaxation (relaxing the scheduling variables $s$ to belong to the interval $[0,1]$ rather than the set $\{0,1\}$). The relaxed set of feasible vectors $s$ will be denoted by $\mathcal S^*$. The relaxed linear program can be written in the form,
\begin{align}
&\arg_s\max\sum_{i,j,f}\theta(i,j,f)s_{ij}^f, \label{relaxedPro1}\\
& s.t\ s_{ij}^f\in[0,1],\ \forall\ i,j,f, \label{relaxedPro2}
\end{align}
where $\theta(i,j,f)=w^fQ_i^f(t)R_{ij}^f$, and $R_{ij}^f=R_{ij}^f(H(t))$ is the rate that is achievable for the link $(i,j)$ if it is transmitting at fixed power, and none of the links it interferes with is on. This is now a separable linear program, and  can then be solved in a distributed fashion.\\
\indent  One algorithm that can be used to solve it in a distributed fashion is the Incremental Gradient Descent  algorithm (IGD) \cite{bertsekas2011incremental}. Let $\mathcal K$ denote the set of all link-flow pairs, i.e., all elements of the form $((i,j),f)$ where $(i,j)\in\mathcal E$ and $f\in\mathcal F$. Then, IGD provides,
\begin{align}
\label{iteraGrad}
s_{n+1}=\Pi_{\mathcal{S}^*}[({s}_n+\alpha v_{k_n}\theta(k_n)s_n],
\end{align}
with $k_n=n\ \text{modulo}\ |\mathcal{K}|+1$, $\alpha$ is a small positive number, $v_{k_n}$ is a vector which is one at its $K_n$-th position and is zero elsewhere, and $\Pi_{\mathcal{S}^*}$ denotes projection onto the  set $\mathcal{S}^*$. Due to the vector $v_{k_n}$, the update of the vector can be performed in a component wise manner. One can perform the update in (\ref{iteraGrad}) in a cyclic manner, going from one element of $\mathcal K$ to the next. At each node, we can do the increment step in (\ref{iteraGrad}) for all the links that originate at that node, and then move to a neighbour. This process then continues cyclically. Thus, we can peform the optimization (\ref{relaxedPro1})-(\ref{relaxedPro2}) in a distributed manner, with messages passed between neighbouring nodes.\\
\indent Since the power of transmission is fixed, and we assume that the channel gains take values from a bounded set, it follows that the rates are bounded by some $\bar R$. Further assume that the weights $w^f$ are bounded by some $\bar w\in\mathbb R$. Let us define,
\begin{align}
F(s)=\sum_{k\in\mathcal K}\theta(k)s(k),\ s\in \mathcal S^*.
\end{align}
 Then, the following result from \cite{bertsekas2011incremental} holds.
\newtheorem{lemma}{Lemma}
\begin{lemma}
	The iterates $\{{s}_n,n\geq 1\}$ given by  (\ref{iteraGrad}) satisfy,
	\begin{align*}
	\lim_{n\to\infty}\sup F({s}_n) \geq \max_{s\in\mathcal S^*}F(s)-C,
	\end{align*}
	where $C=\frac{\alpha\bar w^2\bar R^2|\mathcal{K}|(4|\mathcal{K}|+1) }{2}$.
\end{lemma}
Thus, we can choose $\alpha$ small enough to come close to the optimal value. Note that the algorithm does not require that the age at the destination be available at every node having that flow for computing the optimization. It is only necessary that it be known whether the age exceeds a threshold or not. We can have mini slots at the beginning of each slot, during which the destination node can broadcast a signal at a fixed power, to indicate whether the age has exceeded a threshold. Absence of the signal would indicate that the age is below the threshold. Using this simple signalling scheme, the one bit information corresponding to each flow can be broadcast. 
\section{Simulation Results and Discussion}
We compare the proposed policy, SDSPD, with five other policies. First, we have Backpressure with Dropping (BP-D), which is the same as SDSPD, except that the optimization (\ref{OptRule}), we replace $Q_i^f(t)$ by $Q_{ij}=\max_f(Q_i-Q_j)^+$, and $w^f\equiv 1$. This can be considered as a maxweight (backpressure) policy with dropping. There are two other variants of the SDSPD policy, which use the same scheduling rule as SDSPD, but they do not drop packets. The first of these is SDSPnD-FCFS, which has the FCFS service discipline, and the second, SDSPnD-LCFS, has LCFS service. We also compare with BP-LCFS and BP-FCFS. which are backpresssure  policies without dropping packets, with LCFS and FCFS service respectively. Finally we have  the randomized scheduling policy of \cite{ModianoDist}, which is a randomized stationary policy. It solves an optimization to obtain activation probabilities for links. It does not use instantaneous state information. Comparing with all these schemes allows us to evaluate the performance of the SDSPD algorithm against common scheduling schemes, some of which have been shown to perform well in terms of age.\\
\indent  We consider two example networks. All simulations are run for $10^4$ time slots, and averaged over $100$ such trials. For a theoretical comparison, we use the following lower bound.
\subsection{An Approximate Lower Bound for Age}
Consider a discrete time queue, with a Bernoulli arrival process, so that in each slot, a packet arrives with probability $p$, and  with probability $1-p$, no packet arrives. Let $X$ denote the time between two packet arrivals. Clearly,
\begin{align}
\mathbb EX=\frac{1}{p},\ \mathbb EX^2=\frac{2-p}{p^2}.
\end{align}
  The average age of the arrival process will be,
\begin{align}
\bar{\alpha}=\frac{\mathbb EX^2}{2\mathbb E X}=\frac{2-p}{2p}.
\end{align}
If we assume that the channel takes values $0$ or $1$ with probability $1-q$ and $q$ respectively, the mean time between two time slots in which the channel state is $1$, is $\frac{1}{q}$, and this adds to the average age. Across a system of $n$ such links, we can obtain a lower bound on average age as,
\begin{align}
\frac{2-p}{2p}+\frac{n}{q}. \label{LowerBd}
\end{align}
Observe that this is a loose bound, because it assumes that there is only one flow in the system. In a system with multiple flows, we may be far away from this lower bound.
\subsection{Example Network 1}
\begin{figure}
 	\centering
 	\setlength{\unitlength}{1cm}
 	\thicklines
 	\begin{tikzpicture}[scale=1, transform shape]
	\node[draw,shape=circle, fill={rgb:orange,1;yellow,0;pink,2;green,0}, scale=0.6, transform shape] (v2) at (0,0) {\Huge $2$};
	\node[draw,shape=circle, fill={rgb:orange,1;yellow,0;pink,2;green,0}, scale=0.6, transform shape] (v6) at (0,1.5) {\Huge $6$};
	\node[draw,shape=circle, fill={rgb:orange,1;yellow,0;pink,2;green,0}, scale=0.6, transform shape] (v9) at (3,2) {\Huge $9$};
	\node[draw,shape=circle, fill={rgb:orange,1;yellow,0;pink,2;green,0}, scale=0.6, transform shape] (v3) at (2.5,0) {\Huge $3$};
	\node[draw,shape=circle, fill={rgb:orange,1;yellow,0;pink,2;green,0}, scale=0.6, transform shape] (v11) at (-2,2) {\Huge $11$};
	\node[draw,shape=circle, fill={rgb:orange,1;yellow,0;pink,2;green,0}, scale=0.6, transform shape] (v1) at (-2,0) {\Huge $1$};
	\node[draw,shape=circle, fill={rgb:orange,1;yellow,0;pink,2;green,0}, scale=0.6, transform shape] (v8) at (-2,-2) {\Huge $8$};
	\node[draw,shape=circle, fill={rgb:orange,1;yellow,0;pink,2;green,0}, scale=0.6, transform shape] (v10) at (4,3) {\Huge $10$};
	\node[draw,shape=circle, fill={rgb:orange,1;yellow,0;pink,2;green,0}, scale=0.6, transform shape] (v4) at (3.5,-1) {\Huge $4$};
	\node[draw,shape=circle, fill={rgb:orange,1;yellow,0;pink,2;green,0}, scale=0.6, transform shape] (v5) at (4.5,1) {\Huge $5$};
	\node[draw,shape=circle, fill={rgb:orange,1;yellow,0;pink,2;green,0}, scale=0.6, transform shape] (v7) at (5,-1) {\Huge $7$};
	\draw[line width=0.8mm, dashed] (v2) -- (v6) (v6) -- (v9) (v9) -- (v3) (v3) -- (v2) (v6) -- (v11);
	\draw[line width=0.8mm, dashed] (v2) -- (v1) (v2) -- (v8) (v9) -- (v10) (v3) -- (v4) (v4) -- (v5) (v4) -- (v7);
 	\end{tikzpicture}
 	\caption{Example network 1.}
 	\label{figExampleOne}
 \end{figure}
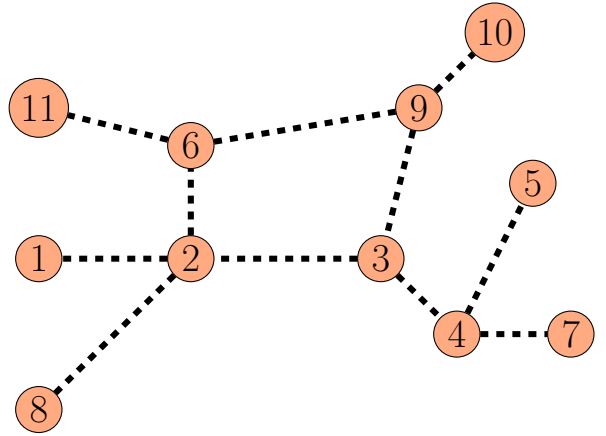

The network considered in this example is given by Figure \ref{figExampleOne}. 
The channel gains take value $0$ or $1$ with probability $0.5$, in each slot. We will assume that if channel gain equals $1$, exactly one packet can be successfully transmitted. This models a situation where the channel is above a threshold with probability $0.5$, and hence ensures succesful transmission. The flows are from node $1$ to $5$ (path $1\to 2\to 3\to 4\to 5$), from node $6$ to node $7$ (path $6\to 2\to 3\to 4\to 7$), from node $8$ to $10$ (path $8\to 2\to 3\to 9\to 10$), from node $11$ to $9$ (path $11\to 6\to 9$) and from node $11$ to node $2$ (path $11\to 6\to 2$). The interference model assumes that any two links that have a common node interfere, and therefore cannot be active simultaneously. All weights $w^f$ in the optimization (\ref{OptRule}) are identically set to one (by choosing $\bar\alpha^f=\infty$ for all $f$). The arrival process is i.i.d Bernoulli across slots, with packet arrival rate $0.1$ for all the flows.\\
\indent  Table \ref{tableOne} gives the value of average AoI obtained at the destination for each flow, for SDSPD, SDSPnD-FCFS, SDSPnD-LCFS, BP-D, BP-FCFS, BP-LCFS and the stationary policy of \cite{ModianoDist}, as well as the loose lower bound $(\ref{LowerBd})$.  
\begin{table}[h]
	\caption{Average AoI for different flows under different policies, for the network in figure \ref{figExampleOne}, with arrival rates of all flows fixed at $0.1$.}
	\label{tableOne}
\begin{tabular}{|p{2cm}|p{0.8cm}|p{0.8cm}|p{0.8cm}|p{0.8cm}|p{0.8cm}|}
\hline
 \ & Flow 1$\to$5 & Flow 6$\to$7 & Flow 8$\to$10 & Flow 11$\to$9 & Flow 11$\to$2\\
 \hline
 Lower Bound & 17.5 & 17.5  & 17.5 & 13.5  & 13.5 \\
\hline
SDSPD & 22.2 & 20.1 & 19.2 & 14.6 & 17.4\\
\hline
BP-D & 24.6 & 20.5 & 19.6 & 14.8 & 17.9\\
\hline
SDSPnD-LCFS & 25.5 & 24.6 & 22.8 & 15.6 & 18.9\\
\hline
BP-LCFS & 37.4 & 31.9 & 27.6 & 16.3 & 23.5\\
\hline
SDSPnD-FCFS & 33.9 & 30.5 & 26.2 & 15.9 & 21.9\\
\hline
BP-FCFS & 47.2 & 37.3 & 30.1 & 16.3 & 25.4\\
\hline
Policy of \cite{ModianoDist} & 190.2 & 242.8 & 149.5 & 61.65 & 112.75\\
\hline
\end{tabular}
\end{table}
It is easy to see that SDSPD is the best performing, and improves over the LCFS policy as well. The FCFS policy performs decently, but the age performance of the FCFS policy will deteriorate as we increase the arrival rates. The stationary policy of \cite{ModianoDist} performs an order worse than the other three, because it does not take into account channel or buffer state information. For SDSPD, the flows also have ages close to the lower bound. Recall that the lower bound was assuming a single flow using up all the resources. Even with five flows in the network, SDSPD performs quite close to the lower bound. The BP-D policy performs close to SDSPD. However, SDSPD offers a slight improvement over BP-D, especially for the first flow.\\
\indent We repeated the simulation for arrival rate $0.13$ for all the flows (see  Table \ref{tableThree}). Here we see that the age performances of the non-dropping policies begin to deteriorate, owing to congestion. The SDSPD and BP-D policies perform well. The age of all the flows of the SDSPD system have reduced, when compared to Table \ref{tableOne}. The policy is able to utilize the higher rate of updates to reduce the overall age. 
\begin{table}[h]
	\caption{Average AoI for different flows under different policies, for the network in figure \ref{figExampleOne}, with arrival rates of all flows fixed at $0.13$.}
	\label{tableThree}
\begin{tabular}{|p{2cm}|p{0.8cm}|p{0.8cm}|p{0.8cm}|p{0.8cm}|p{0.8cm}|}
\hline
 \ & Flow 1$\to$5 & Flow 6$\to$7 & Flow 8$\to$10 & Flow 11$\to$9 & Flow 11$\to$2\\
  \hline
  Lower Bound & 15.2 & 15.2  & 15.2 & 11.2  & 11.2 \\
\hline
SDSPD & 21.2 & 18.4 & 17.3 & 12.5 & 16.2\\
\hline
BP-D & 24.9 & 19.2 & 17.9 & 12.7 & 16.9\\
\hline
SDSPnD-LCFS & 43.1 & 51.6 & 40.4 & 16.2 & 19.9\\
\hline
BP-LCFS & 95.5 & 98.3 & 79.6 & 19.2 & 51.1\\
\hline
SDSPnD-FCFS & 97.8 & 100.3 & 81.9 & 17.6 & 50.6\\
\hline
BP-FCFS & 160.3 & 154.0 & 121.9 & 20.1 & 78.1\\
\hline
Policy of \cite{ModianoDist} & 186.5  & 250.5 & 163.2  & 62.6 & 111.2\\
\hline
\end{tabular}
\end{table}
\indent From the above two tables, it may seem that the policy of \cite{ModianoDist} has the worst performance. However, this is not true in general. As we increase the arrival rates further, the  average AoI for the non dropping policies begin to blow up as expected, owing to congestion. Table \ref{tableFour} summarizes the average AoI values for the different algorithms when arrival rate is $0.14$. The BP-FCFS algorithm performs the worst.\\
\begin{table}[h]
	\caption{Average AoI for different flows under different policies, for the network in figure \ref{figExampleOne}, with arrival rates of all flows fixed at $0.14$.}
	\label{tableFour}
\begin{tabular}{|p{2cm}|p{0.8cm}|p{0.8cm}|p{0.8cm}|p{0.8cm}|p{0.8cm}|}
\hline
 \ & Flow 1$\to$5 & Flow 6$\to$7 & Flow 8$\to$10 & Flow 11$\to$9 & Flow 11$\to$2\\
  \hline
  Lower Bound & 14.6 & 14.6  & 14.6 & 10.6  & 10.6 \\
\hline
SDSPD & 20.9 & 18.1 & 16.8 & 11.9 & 16.1\\
\hline
BP-D & 25.1 & 18.9 & 17.5 & 12.2 & 16.9\\
\hline
SDSPnD-LCFS & 184.7 & 195.8 & 181.2  & 17.5 & 21.3\\
\hline
BP-LCFS & 251.1 & 259.6 & 234.3 & 21.6 & 132.6\\
\hline
SDSPnD-FCFS & 388.7 & 396.1 & 371.9 & 19.7 & 200.8\\
\hline
BP-FCFS & 408.9 & 409.2 & 368.3 & 23.5 & 247.8\\
\hline
Policy of \cite{ModianoDist} & 199.1  & 264.4 & 163.8  & 65.8 & 102.2\\
\hline
\end{tabular}
\end{table}
The above results demonstrate that the SDSPD policy can give low average AoI, close to the lower bound. Next, we demonstrate how we can use the weights $w^f$ to reduce the average AoI even further. This is done by fixing the $\bar{\alpha^f}$ values in (\ref{weightDefn}). The results are given in Table \ref{tableFive}, for the network in figure \ref{figExampleOne}, with arrival rates of all flows fixed at $0.14$.
\begin{table}[h]
		\caption{Average AoI for different flows under the SDSPD policy, for the network in figure \ref{figExampleOne}, with arrival rates of all flows fixed at $0.14$. First column gives the target age for each flow. A $*$ indicates that the target is set to $\infty$ (i.e., no target).}
		\label{tableFive}
	\begin{tabular}{|p{2cm}|p{0.8cm}|p{0.8cm}|p{0.8cm}|p{0.8cm}|p{0.8cm}|}
		\hline
		Target average age $\bar\alpha^f$ for each flow & Flow 1$\to$5 & Flow 6$\to$7 & Flow 8$\to$10 & Flow 11$\to$9 & Flow 11$\to$2\\
		\hline
		*-*-*-*-* & 20.9 & 18.1 & 16.8 & 11.9 & 16.9\\
		\hline
		18-*-*-*-* & 17.3 & 19.6 & 17.7 & 12.0 & 16.4\\
		\hline
		15-*-*-*-* & 16.7 & 20.3 & 18.0 & 12.0 & 16.6\\
		\hline
		15-*-*-*-11 & 16.7 & 21.6 & 18.3 & 12.7 & 12.3\\
		\hline
		*-16-*-*-12 & 22.2 & 16.6 & 17.8 & 12.9 & 12.8\\
		\hline
	\end{tabular}
\end{table}
The first row gives the values of average AoI without targets. In the second row, we fix a target of $18$ for the first flow, and obtain an average AoI of 17.3. In the next row, we set the target to be  $15$, and obtain an average AoI of $16.7$. Recall from Table \ref{tableFour} that the loose lower bound for AoI assuming that only one flow is present was $14.6$, and therefore $16.7$ is a good value for average AoI. The AoI of other flows is only marginally increased. In the next row, we set targets of $15$ and $11$ for the first and last flows (with lower bounds $14.6$ and $10.6$ respectively), and obtain average AoI values of $16.7$ and $12.3$. In the last row we set targets of $16$ and $12$ for the second and last flows, respectively, and obtain $16.6$ and $12.8$ respectively. Thus, the algorithm can provide close to optimal performance, and can prioritize some flows over others if necessary.
\subsection{Example Network 2}
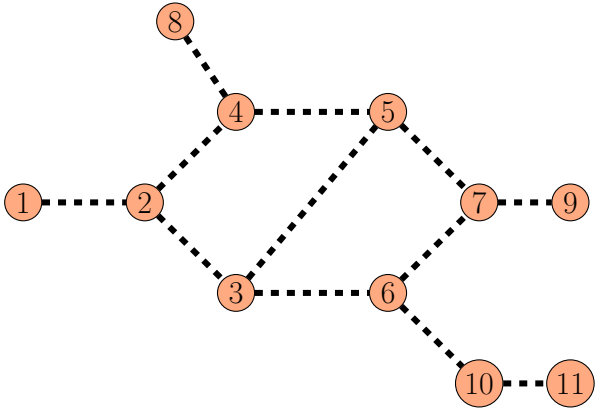
\begin{figure}
 	\centering
 	\setlength{\unitlength}{1cm}
 	\thicklines
 	\begin{tikzpicture}[scale=0.8, transform shape]
	\node[draw,shape=circle, fill={rgb:orange,1;yellow,0;pink,2;green,0}, scale=0.6, transform shape] (v1) at (-2,0) {\Huge $1$};
	\node[draw,shape=circle, fill={rgb:orange,1;yellow,0;pink,2;green,0}, scale=0.6, transform shape] (v2) at (0,0) {\Huge $2$};
	\node[draw,shape=circle, fill={rgb:orange,1;yellow,0;pink,2;green,0}, scale=0.6, transform shape] (v3) at (1.5,-1.5) {\Huge $3$};
	\node[draw,shape=circle, fill={rgb:orange,1;yellow,0;pink,2;green,0}, scale=0.6, transform shape] (v4) at (1.5,1.5) {\Huge $4$};
	\node[draw,shape=circle, fill={rgb:orange,1;yellow,0;pink,2;green,0}, scale=0.6, transform shape] (v5) at (4,1.5) {\Huge $5$};
	\node[draw,shape=circle, fill={rgb:orange,1;yellow,0;pink,2;green,0}, scale=0.6, transform shape] (v6) at (4,-1.5) {\Huge $6$};
	\node[draw,shape=circle, fill={rgb:orange,1;yellow,0;pink,2;green,0}, scale=0.6, transform shape] (v7) at (5.5,0) {\Huge $7$};
	\node[draw,shape=circle, fill={rgb:orange,1;yellow,0;pink,2;green,0}, scale=0.6, transform shape] (v8) at (0.5,3) {\Huge $8$};
	\node[draw,shape=circle, fill={rgb:orange,1;yellow,0;pink,2;green,0}, scale=0.6, transform shape] (v9) at (7,0) {\Huge $9$};
	\node[draw,shape=circle, fill={rgb:orange,1;yellow,0;pink,2;green,0}, scale=0.6, transform shape] (v10) at (5.5,-3) {\Huge $10$};
	\node[draw,shape=circle, fill={rgb:orange,1;yellow,0;pink,2;green,0}, scale=0.6, transform shape] (v11) at (7,-3) {\Huge $11$};
	\draw[line width=0.8mm, dashed] (v1) -- (v2) (v2) -- (v3) (v2) -- (v4) (v4) -- (v5) (v3) --(v6) (v4) -- (v8);
	\draw[line width=0.8mm, dashed] (v5) -- (v7) (v6) -- (v7) (v7) -- (v9) (v6) -- (v10) (v10) -- (v11) (v3) -- (v5);
 	\end{tikzpicture}
 	\caption{Example network 2.}
 	\label{figExampleTwo}
 \end{figure}

The network considered in this example is given in Figure \ref{figExampleTwo}.
The channel, arrival and interference models are the same as in the previous example. The flows are $1\to 2\to 4\to 5\to 7\to 9$, $3\to 2\to 4\to 8$, $4\to 5\to 3\to 6\to 10$ and $4\to 5\to 7\to 6\to 10 \to 11$. Table \ref{tableTwo} and Table \ref{tableSix} depict values of Average AoI for the four flows, under the different policies considered, at arrival rates $0.1$ and $0.13$, respectively. In this set of simulations too, we see that the patterns observed in the previous example hold.
\begin{table}[h]
	\caption{Average AoI for different flows under different policies, for the network in figure \ref{figExampleTwo}, with arrival rates of all flows fixed at $0.1$.}
	\label{tableTwo}
\begin{tabular}{|p{2cm}|p{1cm}|p{1cm}|p{1cm}|p{1cm}|}
\hline
 \ & Flow 1$\to$9 & Flow 3$\to$8 & Flow 4$\to$10 & Flow 4$\to$11\\
 \hline
 Lower Bound & 19.5 & 15.5  & 17.5 & 19.5  \\
\hline
SDSPD & 25.9 & 17.5 & 20.5 & 20.6\\
\hline
BP-D & 29.8 & 17.7 & 21.2 & 21.1\\
\hline
SDSPnD-LCFS & 28.0 & 19.2 & 26.5 & 25.9\\
\hline
BP-LCFS & 42.7 & 21.7 & 27.8 & 26.2 \\
\hline
SDSPnD-FCFS & 37.2 & 20.7 & 27.9 & 27.4 \\
\hline
BP-FCFS & 59.0 & 22.8 & 28.9 & 26.9 \\
\hline
Policy of \cite{ModianoDist} & 238.2 & 104.7 & 185.7 & 209.7\\
\hline
\end{tabular}
\end{table}
\begin{table}[h]
		\caption{Average AoI for different flows under different policies, for the network in figure \ref{figExampleTwo}, with arrival rates of all flows fixed at $0.13$.}
		\label{tableSix}
	\begin{tabular}{|p{2cm}|p{1cm}|p{1cm}|p{1cm}|p{1cm}|}
		\hline
		\ & Flow 1$\to$9 & Flow 3$\to$8 & Flow 4$\to$10 & Flow 4$\to$11\\
		 \hline
		 Lower Bound & 17.2 & 13.2  & 15.2 & 17.2  \\
		\hline
		SDSPD & 25.9 & 15.6 & 18.9 & 18.6\\
		\hline
		BP-D & 32.1 & 15.9 & 20.1 & 19.3\\
		\hline
		SDSPnD-LCFS & 28.8 & 20.3 & 48.5 & 47.2\\
		\hline
		BP-LCFS & 83.1 & 35.4 & 55.4 & 56.1 \\
		\hline
		SDSPnD-FCFS & 79.9 & 32.4 & 55.9 & 58.5 \\
		\hline
		BP-FCFS & 179.6 & 49.5 & 66.0 & 68.4 \\
		\hline
		Policy of \cite{ModianoDist} & 231.9 & 101.7 & 178.7 & 204.7\\
		\hline
	\end{tabular}
\end{table}
\subsection{Discussion}
These experiments seem to suggest that dropping of packets locally at queues can help reduce age. Moreover, we get a policy that is robust to arrival rate variation. Now it may be that in certain applications, it is imperative to get all the packets from the source to the destination, without losing any information. In such cases one may use the SDSPnD-LCFS scheme, which performs the best among all policies without packet dropping. The disadvantage of non-dropping policies, however, is that in case of large arrival rates, the queues will be large, and the time to move all the packets across, from source to destination, will be huge. If the arrival rates are outside the stability region of the policy, this time may very well be not finite. In such a case, it is not even feasible to get all the packets across. Moreover, as the queue lengths build up, the complexity of optimizations used for resource allocation, may also increase. Against all these, SDSPD offers a distinct advantage. Additionally, the dynamically varying weight function allows us to obtain targeted age.
\section{Conclusion and Future Directions}
In this work, we have presented a control policy which reduces the average AoI in a multihop wireless network. The control policy involves dropping of older packets at each queue, in favour of the youngest packet, and using the queue lengths and channel gains at each link. This policy is seen to perform better than policies without dropping, including LCFS schemes. Indeed, in many cases the scheme of dropping packets offers a huge improvement over LCFS schemes. It also performs much better than policies which do not use state information. Further, the average age obtained by the proposed policy is quite close to a theoretical lower bound as well. We further show that we can come even closer to the lower bound by using the age information at the destination. For applications for which there is no need to get all packets across to the destination, dropping of packets in the manner presented can help improve the performance in terms of age. Not keeping a backlog of older packets reduces buffering requirements. Moreover, there is no need to spend energy in transmitting packets which are not fresh. The network capacity is not a bottleneck in the transmission of fresh information. With packet dropping, higher rates of arrivals of packets do not result in an increase in the age due to queueing. We see a monotone decrease in the average age of different flows, as arrival rate increases. This suggests that in systems with packet dropping, the network is no longer a constraint on the optimal sampling rate. Thus, we can fix the sampling rate independent of network considerations, and dependent only on the energy or other requirements of the sampler at the source node.

 Obtaining better theoretical bounds on the age for multihop networks and characterizing age optimal policies would be relevant directions for future research.
\bibliography{survey}
\bibliographystyle{IEEEtran}

\end{document}